\documentclass[aps,prl,twocolumn,floats,amsmath,amssymb,showpacs,floatfix,longbibliography]{revtex4-1}

\usepackage[dvipsnames,rgb,dvips]{xcolor}

\usepackage{graphicx,overpic}
\usepackage{dcolumn}
\usepackage{psfrag}
\usepackage{siunitx}

\renewcommand{\cite}{\citep}
\newcommand{\ve}[1]{\ensuremath{\mbox{\boldmath$#1$}}}
\newcommand{\ma}[1]{\ensuremath{\mathbb{#1}}}
\newcommand{\T}{^{\sf T}}
\newcommand{\gghat}{\ensuremath{\hat{\ve g}}}

\newcommand{\rd}{{\rm d}}

\newcommand{\figlab}[1]{\label{fig:#1}}

\newcommand{\Eqnref}[1]{Eq.~(\ref{eq:#1})}
\newcommand{\Figref}[1]{Fig.~\ref{fig:#1}}

\DeclareMathOperator{\tr}{Tr}
\DeclareMathOperator{\ku}{Ku}
\DeclareMathOperator{\st}{St}

\DeclareMathOperator{\G}{G}

\newcommand{\markerBB}[1]{\protect\includegraphics[width=2mm,clip]{markBB#1.pdf}}
\newcommand{\markerBW}[1]{\protect\includegraphics[width=2mm,clip]{markBW#1.pdf}}

\newcommand{\obs}[1]{{{#1}}}

\begin{document}
\title{Statistical model for the orientation of non-spherical particles settling in turbulence}
\author{K. Gustavsson$^1$, J. Jucha$^{2,\,3}$, A. Naso$^4$, E. L\'ev\^eque$^4$, A. Pumir$^2$ and B. Mehlig$^1$}
\affiliation{$^1$ Department of Physics, Gothenburg University, 41296 Gothenburg, Sweden \\
$^2$ Laboratoire de Physique, Ecole Normale Sup\'erieure de Lyon and CNRS, F-69007 Lyon, France\\
$^3$ Projekttr\"ager J\"ulich, Forschungszentrum J\"ulich GmbH, D-52425 Germany \\
$^4$ LMFA, Ecole Centrale de Lyon and CNRS, F-69134 Ecully, France}

\begin{abstract}
The orientation of small anisotropic particles settling in a turbulent fluid determines some essential properties of the suspension.  We show that the orientation distribution of small heavy spheroids settling through turbulence can be accurately predicted by a 
simple Gaussian statistical model that takes into account particle inertia and  provides a quantitative understanding of the orientation distribution on the problem parameters when fluid inertia is negligible. Our results open the way to a parameterisation of the distribution of ice-crystals in clouds, and potentially leads to an improved understanding of radiation reflection, or particle aggregation through collisions in clouds.
\end{abstract}

\maketitle

How non-spherical objects settle in a turbulent environment is a
highly relevant question in several domains. An example is provided by
very small ice crystals in clouds (size $\sim\SI{100}{\micro\metre}$), which grow
through aggregation to form precipitation size particles (size $\sim\SI{1}{\milli\metre}$)~\cite{Pru78,Cho81,ChenLamb94,Hub14}. The settling of plankton in the
ocean \cite{Rui2004,Cen13,Berglund2016} can induce patchiness of the population,
therefore affecting mating, feeding and predation \cite{Gua12}. In these
problems, the orientational degrees of freedom clearly affect not only
settling and collision properties, but also light reflection \cite{Yang15}.
As a prerequisite to a description of these effects,
this Letter provides an understanding of the orientation
statistics of small spheroids settling in a turbulent environment
based on a statistical model,
under the assumption that fluid inertia can be neglected.

The interaction between turbulence and settling leads to intriguing
phenomena, even in the simpler case of spherical particles. Maxey found that
 turbulence increases the settling speed of a single small particle \cite{Max87,Goo14}.
Substantial progress was recently achieved in understanding
how two spherical particles settling together move relative to each other and collide ~\cite{Gus14e,Bec14,Ireland,Mathai2016,Parishani}.

In a fluid at rest the orientation dynamics of slowly settling {\em non-spherical} particles is determined by
weak torques resulting from fluid inertia~\cite{Kha89,Sub15,Can16,Koc16}.
Turbulence affects the orientation of such particles through
turbulent vorticity and strain. In the absence of settling this is well understood \cite{Jef22,Pum11,Par12,Che13,Gus14,Byr15,Zha15,Voth15,Voth16,Berglund2016}.
Neglecting fluid inertia, the direct numerical simulations (DNS) of turbulence by
Siewert \textit{et al.} \cite{Siew14a} demonstrated
that settling induces a bias in the orientation
distribution of the particles. The physical origin of this bias
is not known, and it is not understood how the bias depends on the parameters of the problem: the turbulent
Reynolds number, ${\rm Re}_\lambda$, the Stokes number
(particle inertia), the gravitational acceleration, and the particle shape.
Also, how significant are non-Gaussian, intermittent small-scale features of
the turbulent flow \cite{Sch14}, such as intense vortex tubes \cite{Cho81} in aligning the particles?

\begin{figure}[t]
\begin{overpic}[width=1.0\columnwidth]{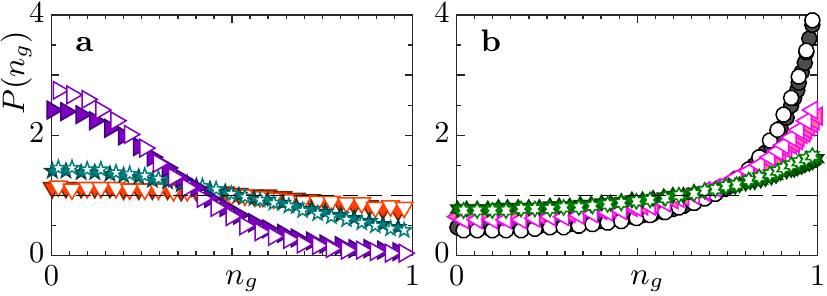}
\end{overpic}
\vspace{-0.5cm}
\caption{\label{fig:comparison}
Orientational bias of spheroids settling in turbulence.
Distribution $P(n_g)$ of $n_g\equiv|\ve n\cdot\gghat|$; particle symmetry-vector $\ve n$ and direction $\gghat$ of gravity.
{\bf a} DNS results for $P(n_g)$ for oblate spheroids (aspect ratios $\lambda = 0.01$ (\markerBB{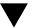}), $0.02$ (\markerBB{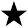}), $0.05$ (\markerBB{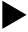}).
Statistical-model simulations, see text, open symbols.
Dashed line shows isotropic distribution $P(n_g)=1$.
{\bf b} Same, but for prolate spheroids: $\lambda = 5$ (\markerBB{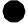}), $7.5$ (\markerBB{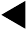}), $10$ (\markerBB{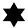}).
The parameters chosen: ${\rm Re}_\lambda\approx 95$, $F_{\rm K}\approx 70$, ${\rm St}_{\rm K}\approx 4\min(\lambda,1/\lambda)$
correspond to values relevant to cloud physics, see \obs{Supplemental Material}~\cite{SM}. }
\end{figure}

To answer these questions we analyse a statistical model for the orientation of small heavy spheroids settling in homogeneous isotropic turbulence, for parameters relevant to cloud physics, and compare with
results based on DNS of turbulence. Fig.~\ref{fig:comparison} shows the predicted bias in the distribution of the vector $\ve n$ pointing   along the particle symmetry axis.
The statistical-model predictions agree very well with the DNS results.
This shows that that non-Gaussian turbulent fluctuations are not important.
The statistical model explains the sensitive parameter dependence of the DNS results. This is important because it allows us to parameterise the bias, to quantitatively understand the physical properties of the system.
We analyse the model by an expansion in the \lq Kubo number\rq{} ${\rm Ku}$, a dimensionless correlation time of the flow \cite{Gus16}.  Pad\'e-Borel resummation yields excellent agreement with numerical simulations at $\ku=0.1$, and qualitative agreement with DNS of turbulence.  At larger $\ku$ the theory fails to converge, but the model still explains qualitatively the underlying mechanisms. Last, we discuss possible effects of fluid inertia.

{\em Formulation of the problem.}
The equations of motion for translation and rotation of a particle reads
\begin{equation}
\label{eq:eom}
m \ddot{\ve x} = \ve f + mg\gghat\,,\quad \dot {\ve n}
= \ve \omega \wedge \ve n\,, \quad
 \tfrac{\rd}{\rd t} \big(\ma J(\ve n) \ve \omega\big)  = \ve T\,.
\end{equation}
Here $g$ is the gravitational acceleration (direction $\gghat$), $\ve x$ is the
position of the particle, $\ve n$ its symmetry vector, $m$ its mass, $\ve \omega$ its angular velocity,
and $\ma J(\ve n)$ is its inertia tensor in the lab frame.
In the point-particle approximation,  force $\ve f$ and torque  $\ve T$ on a spheroid are \cite{Kim:2005,Marchioli2010,Gus14}:
\begin{align}\label{eq:resistance}
\begin{bmatrix}
\ve f \\
\ve T
\end{bmatrix}
= m\gamma
\begin{bmatrix}
\ma M^{(t)}& 0 &0 \\
0 & \ma M^{(r1)} & \ma M^{(r2)}
\end{bmatrix}
\begin{bmatrix}
\ve u - \ve v \\
\ve \Omega-\ve \omega \\
\ma S
\end{bmatrix}.
\end{align}
In Eq.~(\ref{eq:resistance}), $\ve v$ is the particle velocity, $\ve u(\ve x,t)$ is the turbulent velocity field,
$\ve \Omega\equiv\tfrac{1}{2}\ve \nabla \wedge \ve u$
is half the turbulent vorticity, $\ma S$ is the strain-rate matrix,
the symmetric part of the matrix $\ma A$ of fluid-velocity gradients
(its antisymmetric part is called $\ma O$),
and $\ma M$ are translational and rotational resistance tensors:
$\ma M^{(t)}\equiv C^{(t)}_\perp\ma I+(C^{(t)}_\parallel\!-\!C^{(t)}_\perp)\ve n\ve n^{\sf T}$,
$\ma M^{(r1)}\equiv {K}^{{(r1)}}_\perp\ma I+({K}^{{(r1)}}_\parallel\!-\!{K}^{{(r1)}}_\perp)\ve n\ve n^{\sf T}$,
and $\ma M^{(r2)}$ is a third-rank tensor.
For a fore-aft symmetric particle, the equations of motion (\ref{eq:eom},\ref{eq:resistance}) are invariant under  $\ve n\to-\ve n$, so that only the magnitude $n_g\equiv |\ve n \cdot \gghat|$ can play a role in the dynamics.
The form of $\ma M^{(r2)}$ and of the $C$ and $K$-coefficients
are known for spheroidal particles, see Supplementa\obs{l} Material (SM)~\cite{SM} and Ref.~\cite{Fries16}.
The parameter $\gamma\equiv 9\nu\rho_f/(2{a_\parallel a_\perp}\rho_p)$
is Stokes constant,  $\nu$ is the kinematic viscosity of the fluid, $\rho_f$ and $\rho_p$ are fluid and particle mass densities, $2a_\parallel$ \obs{is the} length of the particle symmetry axis, and $2a_\perp$ is 
\obs{the particle} diameter.

Our DNS of turbulence use the code described in \cite{Vos13} and in \obs{the} SM~\cite{SM}.
The Kolmogorov scales $u_{\rm K}$, $\eta_{\rm K}$, and $\tau_{\rm K}$
are determined by the dissipation rate $\varepsilon \equiv\nu \langle\tr\ma A\ma A\T\rangle$  (the average is along steady-state Lagrangian trajectories), and by $\nu\approx $\SI{1e-5}{\metre\squared\per\second} (air).
The particle aspect ratio is $\lambda \equiv a_\parallel/a_\perp$.
The simulations were done for spheroids of varying $\lambda$ and \obs{with} ${{\rm max}(a_\parallel,a_\perp)} = $\SI{150}{\micro\metre}, much smaller than $\eta_{\rm K}$ for
values of $\varepsilon$ pertaining to mixed-phase clouds (DNS: $\varepsilon \! \approx\! 1,\, 16,$ and $\SI{256}{\centi\metre\squared\per\second\cubed}$).
Particle inertia is measured by the Stokes number
$\st_{\rm K} \equiv (\gamma \tau_{\rm K})^{-1}$.
The mass-density ratio is $\obs{\rho_p/\rho_f}\approx 1000$ (ice crystals in air), and the \obs{dimensionless} gravity parameter \obs{is defined as} ${\rm F}_{\rm K} \equiv g \tau_{\rm K}/u_{\rm K}$.

{\em Statistical model.} The model is appropriate for particles smaller than $\eta_{\rm K}$. We approximate the universal \cite{Sch14} dissipative-range turbulent fluctuations by  an incompressible, homogeneous, isotropic Gaussian random velocity field $\ve u(\ve x,t)$ with zero mean,  correlation length $\ell$, correlation time $\tau$, and rms speed $u_0$ \cite{Gus16} (details given in the SM \cite{SM}).
In the persistent limit \cite{Gus16}, for $\ku  \equiv u_0 \tau/\ell > 1$, the model parameters $\st\equiv (\gamma\tau)^{-1}$ and ${\rm F}\equiv g \tau/u_0$ map to
$\st_{\rm K}=\sqrt{5}\ku\st$
and $F_{\rm K}=[F/(5\ku)]\ell/\eta_{\rm K}$.
Here $\ell/\eta_{\rm K}$ is the ratio between the size of the dissipation range and the Kolmogorov length. In turbulence this ratio
depends weakly on the Reynolds number ${\rm Re}_\lambda$ \cite{Cal09},
$\ell = c \eta_{\rm K} {\rm Re}_\lambda^{1/2}$. For the data shown in Fig.~\ref{fig:comparison} we have
${\rm Re}_\lambda = 95$, \obs{and} Fig.~S1 in SM \cite{SM} shows results for other values of ${\rm Re}_\lambda$.
We find good agreement between the statistical-model results at large $\ku$ and the DNS for
$c\approx 1.3$.  For $\ku>1$, the model predictions depend on two parameter combinations only~\cite{Gus16}, $\ku \st$ and ${\rm F}/\ku$. In terms of the DNS parameters this means that the orientation bias depends
only on $\st_{\rm K}$ and $F_{\rm K}{\rm Re}_\lambda^{-1/2}$.

{\em Perturbation theory.}
Eqs.~(\ref{eq:eom},\ref{eq:resistance}) are solved
by expansion in powers of $\ku$~\cite{Gus11a,Gus16}.
We outline the essential steps below, details are given in the SM \cite{SM}.
We use dimensionless variables: $t'\!\equiv\!t/\tau, \ve r'\!\equiv\! \ve r/\ell, \ve u'\!\equiv \!\ve u/u_0$, and drop the primes.
To calculate the steady-state distribution of $n_g \equiv |\ve n\cdot \gghat|$
we must evaluate the fluctuations of the fluid-velocity gradients along particle paths.
This is achieved by an expansion in $\delta\ve x_t \equiv \ve x_t -\ve x^{({\rm d})}_t$
around the deterministic solution $\ve x^{({\rm d})}_t$ of Eqs.~(\ref{eq:eom},\ref{eq:resistance})
for $\ve u=0$.
This gives expansions in powers of $\ku$~\cite{Gus16}:
\begin{widetext}
\begin{align}
\label{eq:nt}
\ve n_t&=\ve n_0 + \ku\int_0^t\!\!\!{\rm d}t_1\,[1-{\rm e}^{(t_1-t)C^{{(r1)}}_\perp/\st}]
\left.(\delta\ve x_{t_1}\cdot\ve\nabla)\big\{\ma O(\ve x,t_1)\ve n_0
+{\tfrac{\lambda^2-1}{\lambda^2+1}}\big[\ma S(\ve x,t_1)\ve n_0
- \big(\ve n_0\cdot \ma S(\ve x,t_1) \ve n_0\big)\ve n_0\big]\big\}\right|_{\ve x=\ve x_{\scriptstyle t_1}^{(\rm d)}} \nonumber \\[-2mm]
&+\ku\int_0^t{\rm d}t_1(1-{\rm e}^{C^{{(r1)}}_\perp(t_1-t)/\st})\big[\ma O(t_1)\ve n_0+
\tfrac{\lambda^2-1}{\lambda^2+1}\big(\ma S(t_1)\ve n_0 - (\ve n_0\T\ma S(t_1)\ve n_0\big)\ve n_0)\big]   \\[-1mm]
&+\ku^2\int_0^t{\rm d}t_1\int_0^{t_1}{\rm d}t_2
 \big[ \ve c^{(OO)}_{ijkl}(\ve n_0;t,t_1,t_2) O_{ij}(t_1)O_{kl}(t_2)+\mbox{similar terms with $OS$ and $SS$}\big ]\,.
\nonumber
\end{align}
\end{widetext}
The matrices $\ma O$ and $\ma S$ are evaluated along deterministic paths
$\ve x_t^{({\rm d})}=\ve x_0 + \ve v_s(\ve n_0)\, t$ (Fig.~\ref{fig:alignment_smallKu}{\bf a}) with
settling velocity
\begin{align}
\label{eq:vs}
\ve v_s(\ve n_0)
& ={\rm F}\,\st \Big[\frac{\ma I}{C^{(t)}_\perp}\!+\!\frac{\ve n_0\ve n_0\T}{C^{(t)}_\parallel}\!-\!\frac{\ve n_0\ve n_0\T}{C^{(t)}_\perp}\Big] \cdot \gghat\,.
\end{align}
Also, $C^{(r1)}_\perp\equiv 5\lambda K^{(r1)}_\perp/(a_\parallel a_\perp(\lambda^2+1))$.
Eq.~(\ref{eq:vs}) is the lowest-order solution of Eqs.~(\ref{eq:eom},\ref{eq:resistance}).
The terms in \Eqnref{nt} that do not involve $\delta\ve x_t$ depend only on the
history of the fluid-velocity gradients along the paths $\ve x_t^{({\rm d})}$ (\lq history contribution\rq{}).
The $\ve c$-coefficients contain at most five powers of $\ve n_0$,
and one must sum over all tensor products allowed by symmetry (Einstein convention).
See SM \cite{SM}.

The first integral shown in Eq.~(\ref{eq:nt}), by contrast,
depends on $\delta\ve x_t$. It is therefore
sensitive to how turbulence modifies
the settling paths (\lq preferential sampling\rq{} \cite{Gus16}).

We determine the steady-state moments $\langle (\ve n_t\cdot \gghat)^p\rangle_\infty$ by first calculating the moments conditional on the initial orientation $\ve n_0$, using Eq.~(\ref{eq:nt}) and the relation
\begin{eqnarray}
\label{eq:mts}
&&\langle(\ve n_t\cdot\gghat)^{p}\rangle_{\raisebox{-0.1mm}{$\scriptstyle \ve n_0$}}\!\!\!=\!
(\ve n_0\cdot\gghat)^{p}\!+\!p\ku(\ve n_0\cdot\gghat)^{p-1}\langle \ve n_t^{(1)}\!\cdot\gghat\rangle_{\raisebox{-0.1mm}{$\scriptstyle \ve n_0$}}\\
&&\!+\!\tfrac{p}{2}\ku^2(\ve n_0\cdot\gghat)^{p-2}\big\langle 2(\ve n_0\cdot\gghat)(\ve n_t^{(2)}\!\cdot\gghat) \!+\! (p\!-\!1)(\ve n_t^{(1)}\!\cdot\gghat)^2\big\rangle_{\raisebox{-0.1mm}{$\scriptstyle \ve n_0$}}\nonumber
\end{eqnarray}
where
$\ve n_t^{(i)}$ is the coefficient of $\ku^i$ in \Eqnref{nt}.
Eq.~(\ref{eq:mts}) is valid to order $\ku^2$.
We average over the fluid-velocity fluctuations as described in Ref.~\cite{Gus16}.
The moments are independent of the initial position $\ve x_0$ due to homogeneity of the flow.
We expect that effects of the initial velocity $\ve v_0$ and angular
velocity $\ve \omega_0$ decay exponentially, so that they do not affect the steady state. We therefore
set both to zero. {Only the $\ve n_0$-dependence matters.}
In this way we obtain
expressions for $\langle(\ve n_t\cdot\gghat)^{p}\rangle_{\raisebox{-.1mm}{$\scriptstyle \ve n_0$}}$, which involve
secular terms that increase linearly with time as $t\to\infty$.
But these terms must vanish since $\ve n_t$ is a unit vector. This condition yields
a recursion relation for the steady-state averages $\langle(\ve n\cdot\ve\gghat)^p\rangle_\infty$, independent of $\ve n_0$.
This recursion is valid for arbitrary values of
$\G \equiv\ku\,{\rm F}\st/C^{(t)}_\perp$, and to order $\ku^0$.
Note that $\G$ can be large even if $\ku$ is small.
We solve the recursion by a series expansion in small $\G$:
\begin{align}
\label{eq:moments}
\langle(\ve n\cdot\gghat)^{2p}\rangle_\infty\!=\!
\frac{1}{2p\!+\!1}\!+\!
\sum_{i=1}^\infty\frac{\G^{2i}\sum_{j=1}^i p^jA^{(2i)}_{j}(\st,\lambda)}{\prod_{k=1}^{i+1}(2 p + 2 k - 1)}\,.
\end{align}
The coefficients $A^{(2i)}_{j}(\st,\lambda)$ depend on the shape and inertia of the particle, but not on $\G$ or $p$.
From Eq. (\ref{eq:moments}) we obtain the Fourier transform of the probability distribution of $n_g=|\ve n\cdot\gghat|$.
Inverse Fourier transformation yields the distribution. To order $\G^4$ we find:
\begin{eqnarray}
\nonumber
P(n_g) &=& 1 \!+\! \frac{1}{4}(3n_g^2\!-\!1)A^{(2)}_{1}\G^2+\tfrac{1}{32}\big[2(1\!-\!n_g^2)(5n_g^2 \!-\! 1)A^{(4)}_{1}\nonumber \\
&+& (1 - 18n_g^2 + 25n_g^4)A^{(4)}_{2}\big]\G^4+\dots\,.
\end{eqnarray}
The lowest-order term corresponds to a uniform distribution of $\ve n_t$.
Let us examine the $\G^2$-term. It turns out that $A_1^{(2)}$ is negative for disks and positive for rods (see
Fig.~S2 in \obs{the} SM \cite{SM}).
This explains that the orientation of settling disks is biased: disks tend to fall edge on and rods settle tip first (as in Fig.~\ref{fig:comparison}).
\begin{figure}[t]
\begin{overpic}[width=\columnwidth]{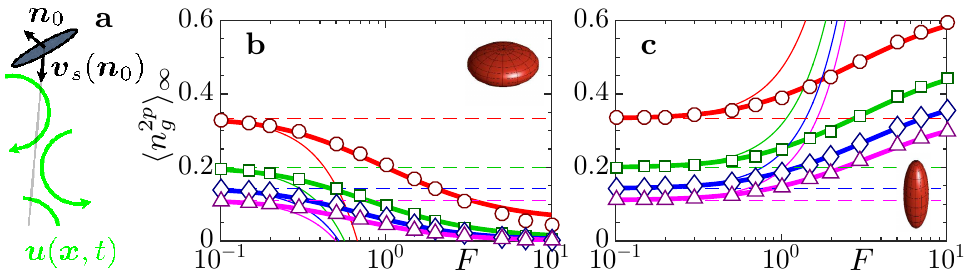}
\end{overpic}
\caption{\figlab{alignment_smallKu}
 {\bf a} Deterministic settling path along $\ve v_s(\ve n_0)$, \obs{independent of} the instantaneous fluid velocity $\ve u(\ve x,t)$.  {\bf b}, {\bf c} Moments of $n_g=|\ve n\cdot\gghat|$, for $p=1$ (\markerBW{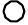}), $p=2$ (\markerBW{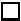}), $p=3$ (\markerBW{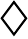}), $p=4$ (\markerBW{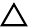}).  Dashed lines: moments of isotropic orientation distribution.  Thin solid lines:  \Eqnref{moments} to $O(\G^2)$.  Thick solid lines:  order $8$-by-$8$ Pad\'e-Borel resummation of \Eqnref{moments} to $O({\rm G}^{32})$.  Parameters: $\ku=0.1$, $\st=10$, $\lambda^2=0.1$ ({\bf b}) and $\lambda^2=10$~({\bf c}). }
\end{figure}
\begin{figure*}
\begin{overpic}[width=0.91\textwidth]{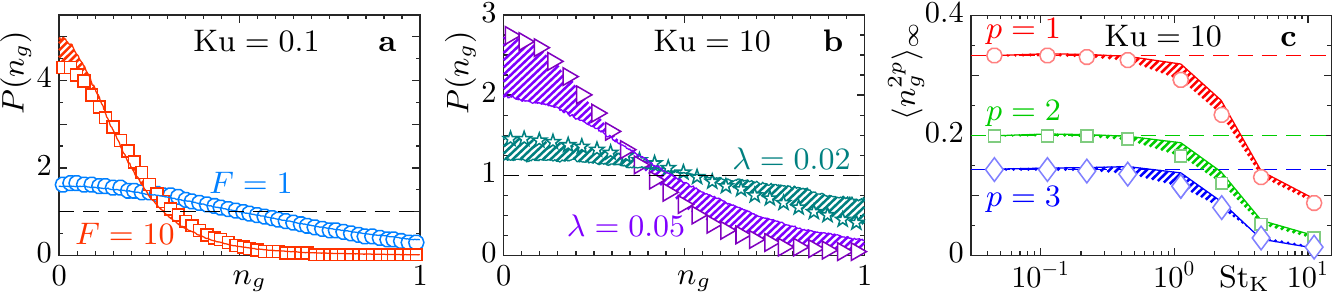}
\end{overpic}
\caption{\label{fig:alignment}
History effect causes orientation bias.
{\bf a} Distribution of $n_g$ based on full statistical-model simulations (symbols), and based on straight deterministic paths (see Fig.~\ref{fig:alignment_smallKu}{\bf a}), solid lines. Parameters: $\lambda=1/\sqrt{10}$, $\ku=0.1$, $\st=10$, $F=1$ (\markerBW{1}), ${\rm F}=10$ (\markerBW{2}).
{\bf b} Same for parameters corresponding to \obs{the} data in Fig.~\ref{fig:comparison}{\bf a}, $\lambda=0.02$ (\markerBW{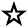}), $\lambda=0.05$ (\markerBW{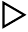}).
 Preferential-sampling contribution is hatched.
{\bf c} Moments $\langle n_g^{2p}\rangle_\infty$ from
statistical-model simulations in the persistent limit ($\ku=10$) against $\st_{\rm K}$ ($p=1$, \markerBW{1}; $p=2$, \markerBW{2}; $p=3$, \markerBW{3}). Parameters $\lambda=1/\sqrt{10}$, ${\rm F}_{\rm K}\approx 2.5$. Also shown are simulations based
on straight deterministic paths (solid lines).} 
\end{figure*}

{\em Pad\'e-Borel resummation.}
Now consider higher orders in the $\G$-expansion. The series (\ref{eq:moments}) is asymptotically
divergent and must be resummed. Fig.~\ref{fig:alignment_smallKu} demonstrates that Pad\'e-Borel resummation
\cite{Ben78,Gus16} of the series yields excellent results. Shown are results from a resummation of (\ref{eq:moments}) to order $\G^{34}$ (thick solid lines).
These results agree very well with numerical simulations of the statistical model for $\ku=0.1$ and $\st=10$ (symbols).
The resummed theory works up to ${\rm G}=10$, and in this range
the bias increases with increasing ${\rm G}$.
The resummed theory also predicts that the moments increase
as $\st$ increases, for fixed $\G$. A more detailed analysis of the recursion leading to Eq.~(\ref{eq:moments}) reveals, however, that the limit ${\rm G}\to\infty$ is delicate.
Perfect alignment requires $\lambda=\infty$ \cite{SM}.

In summary, perturbation theory in $\ku$ shows that turbulence gives rise to an orientation bias
 (Fig.~\ref{fig:alignment_smallKu}),
in excellent agreement with statistical-model simulations at $\ku=0.1$ and in qualitative agreement with
DNS (Fig.~\ref{fig:comparison}).

The calculations leading to Eq.~(\ref{eq:moments}) reveal
that each moment $\langle (\ve n_t\cdot \gghat)^{2p}\rangle_\infty$
is a sum of two contributions that stem from the \lq preferential sampling\rq{} and \lq history\rq{} terms in Eq.~(\ref{eq:nt}).
For small $\ku$ the history effect is dominant, the orientation bias is entirely determined by the history of fluid-velocity gradients along
straight deterministic paths, Fig.~\ref{fig:alignment_smallKu}{\bf a}.
Decomposing the leading-order contribution
as $A_1^{(2)} \!=\! A_{1,\rm pref.}^{(2)}\!+\!A_{1,\rm hist.}^{(2)}$ we find that $|A_{1,\rm pref.}^{(2)}|\ll |A_{1,\rm hist.}^{(2)}|$ (Fig.~{S2}{\bf b} in \obs{the} SM \cite{SM}).
\Figref{alignment}{\bf a} leads to the same conclusion. It shows
 the distribution $P(n_g)$ for $\ku=0.1$. Also shown is $P(n_g)$
computed for particles falling with constant velocity $\ve v=\ve v_{\rm s}(\ve n_0)$.
We choose the squared initial orientation $\ve n_0\ve n_0\T$ in (\ref{eq:vs}) as the steady state average $\langle\ve n_0\ve n_0\T\rangle_\infty$, evaluated using the small-$\ku$ theory.
This corresponds to keeping just the history contribution
to $\langle n_g^{2p}\rangle_\infty$.
We observe excellent agreement with the full statistical-model simulations. This shows that the history effect causes the orientation bias at small \obs{values of} $\ku$.

{\em Persistent limit.} In the persistent limit we use numerical simulations with $\ku=10$ to analyse the orientation bias in the same way as for small $\ku$. The result is shown in \Figref{alignment}{\bf b} (parameters correspond to two curves in Fig.~\ref{fig:comparison}({\bf a}). We plot the full statistical-model distribution and results for particles with a constant velocity~(\ref{eq:vs}) that neglects preferential sampling.  For the data in  \Figref{alignment}{\bf b},
the average $\langle\ve n_0\ve n_0\T\rangle_\infty$  is computed using statistical-model simulations.
We see that the history effect makes a substantial contribution to $P(n_g)$. But since the distributions do not match, we infer that
preferential sampling also contributes. This contribution is hatched in \Figref{alignment}{\bf b}.

{\em Limit of large settling speeds.}
Fig.~\ref{fig:alignment}{\bf c} shows the moments $\langle n_g^{2p}\rangle_\infty$ for $p=1,2,3$ in the persistent limit as functions
of the DNS Stokes number $\st_{\rm K}$.
Open symbols denote full statistical-model simulations, solid lines correspond to simulations based on straight deterministic paths.
At intermediate Stokes numbers we see a clear difference
between the two simulations, preferential sampling is important in this region.

As the Stokes number grows, however, the Figure demonstrates that preferential sampling ceases to play a role.
In this limit the orientation bias is entirely caused by the history effect.
The bias shown in Fig.~\ref{fig:alignment}{\bf c} increases as $\st_{\rm K}$ increases.
But as the perturbation theory indicates, the limit of large $\G$ is quite subtle. Statistical-model simulations for $\ku=1$ show that the degree of alignment starts to decrease for very large $\G$.

{\em Conclusions.}
We analysed a statistical model for the orientational dynamics of small heavy spheroids settling in turbulence.
The predictions of the model agree well with our own numerical results based on DNS of homogeneous isotropic turbulence (Fig.~\ref{fig:comparison}).
Our statistical-model analysis shows that there are  two distinct competing mechanisms causing the orientation bias:
preferential sampling and the history effect. The latter dominates for large settling speeds, but it makes substantial contributions also in other parameter regimes. Preferential sampling dominates only when the bias is negligibly small.
When the bias is significant, the history effect explains at least about 50\% of the \obs{bias} observed in Fig.~\ref{fig:comparison}.

We have shown that the orientation alignment depends on combinations of dimensionless numbers:
$\st_{\rm K}$ and $F_{\rm K}{\rm Re}_\lambda^{-1/2}$.
Our analysis shows that it is the small-scale properties of the flow that determine the orientation alignment.
The ${\rm Re}_\lambda$-dependence arises only because it determines the ratio between the smooth scale $\ell$ to $\eta_{\rm K}$. We note that $F_{\rm K}{\rm Re}_\lambda^{-1/2}$ equals the ratio of
the settling velocity and the rms turbulent velocity fluctuations.

Our results pertain to small ice crystals settling in turbulent clouds, and allow us to model the sensitive dependence of the effect upon particle shape, size, and the turbulence intensity. This is important since turbulent dissipation rates vary widely in clouds.
Our results predict strongly varying degrees of alignment.
That the statistical model is in excellent agreement with the DNS opens a way to parameterise the orientation distribution of ice-crystals in clouds. This potentially leads to an improved understanding of the radiative properties of clouds, and of particle aggregation through collisions in clouds.

The present work is based on the point-particle approximation of heavy particles, which neglects the effect of fluid inertia.
This requires the particle Reynolds number ${\rm Re}_p \equiv a v_c/\nu$ to be small, where $a={\rm max(a_\parallel,a_\perp)}$.  Estimating the slip velocity
$v_c$ by the Stokes settling speed, we find that ${\rm Re}_p$ is of order unity for the data shown in Fig.~\ref{fig:comparison}, so the condition is marginally
{satisfied}.
The shear Reynolds number, ${\rm Re}_s$, must also be small. Since ${\rm Re}_s \equiv a^2 \sqrt{\langle{\rm tr}\,\ma S^2\rangle}/\nu\sim (a/\eta_{\rm K})^2$~\cite{Can16b}, this condition is satisfied for small particles.

Lopez {\em et al.}~\cite{Lop17} analysed the orientational dynamics of rods settling in a vortical flow.
For small ${\rm Re}_p$ they found a bi-modal distribution, with peaks at $n_g=0$ and $1$.
They explain the peak at $n_g=0$ by the effect of fluid inertia. Our \obs{results may} explain the peak at $n_g=1$.
These results, although not for a turbulent flow, indicate that turbulent and fluid-inertia torques compete in general.
How to model this competition is an open question.  For small Stokes numbers one may formulate an {\em ad-hoc} model by simply adding turbulent and fluid-inertia torques, along the lines suggested in Ref.~\cite{Lop17}. But in general it remains a challenge to take into account effects due to fluid inertia from first principles, in a turbulent environment.
Simulations resolving particle and fluid motion~\cite{Homann,Fornari} and experiments \cite{Tra15,variano2016,Voth,Voth2017} for micron-sized particles in turbulence are needed to test the predictions, and to determine the orientational dynamics of larger particles where fluid inertia  must matter \cite{Voth2017}.
Finally, how to extend the ideas developed here to  particles lighter than the fluid remains a challenging task.

\acknowledgments{
{\em Acknowledgments}.
This work was supported by Vetenskapsr\aa{}det [grant number 2013-3992], Formas [grant number 2014-585], and by the grant \lq Bottlenecks for particle growth in turbulent aerosols\rq{} from the Knut and Alice Wallenberg Foundation, Dnr. KAW 2014.0048.
  The numerical computations used resources
 provided by C3SE and SNIC.}


%

\end{document}


\title{Supplemental Material: Angular dynamics of non-spherical particles settling in turbulence}
\author{K. Gustavsson}
\affiliation{Department of Physics, Gothenburg University, 41296 Gothenburg, Sweden}
\author{J. Jucha}
\affiliation{Laboratoire de Physique, Ecole Normale Sup\'erieure de Lyon and CNRS, F-69007, Lyon, France}
\affiliation{Projekttr\"ager J\"ulich, Forschungszentrum J\"ulich GmbH, D-52425, Germany}
\author{A. Naso}
\affiliation{LMFA, Ecole Centrale de Lyon and CNRS, F-69134, Ecully, France}
\author{E. L\'ev\^eque}
\affiliation{LMFA, Ecole Centrale de Lyon and CNRS, F-69134, Ecully, France}
\author{A. Pumir}
\affiliation{Laboratoire de Physique, Ecole Normale Sup\'erieure de Lyon and CNRS, F-69007, Lyon, France}
\author{B. Mehlig}
\affiliation{Department of Physics, Gothenburg University, 41296 Gothenburg, Sweden}
{}
\maketitle
\section{Details on equations of motion}
We begin by recalling the form of the resistance tensors, defined by Eq.~(2)
in the main text.  The tensor ${\ma M}^{(t)}$
relates the force on the particle to the slip velocity, the difference between
the fluid and the particle velocities.  The superscript refers to the translational ($t$) degrees of freedom. The tensor is expressed as:
\begin{equation}
M^{(t)}_{ij} \equiv
C^{(t)}_\perp \delta_{ij} + ( C^{(t)}_\parallel - C^{(t)}_\perp ) n_i n_j\,.
\label{eq:def_Mt}
\end{equation}
For the rotational degrees of freedom, superscript $(r)$, the resistance
tensor is decomposed into two terms, ${\ma M}^{(r1)}$ and  ${\ma M}^{(r2)}$.  The first
term relates the torque to the angular slip velocity, the difference between
the rotation rate of the fluid and that of the particle. The second term
relates the torque to the local strain in the fluid.
The expression for ${\ma M}^{(r1)}$ is similar to Eq.~(\ref{eq:def_Mt}):
\begin{equation}
M^{(r1)}_{ij} \equiv
{K}^{(r1)}_\perp \delta_{ij} + ({K}^{(r1)}_\parallel - {K}^{(r1)}_\perp ) n_i n_j\,.
\label{eq:def_Mr1}
\end{equation}
The expression for ${\ma M}^{(r2)}$ is of the form:
\begin{equation}
M^{(r2)}_{ijk} = {K}^{(r2)}\epsilon_{ijl} n_k n_l\,.
\label{eq:def_Mr2}
\end{equation}
Here $\epsilon_{ijl}$ is the completely anti-symmetric Levi-Civita tensor, and the summation convention is used: repeated indices are summed from $1$ to $3$. For spheroidal particles, the coefficients
in Eqs.~(\ref{eq:def_Mt}), (\ref{eq:def_Mr1}), and (\ref{eq:def_Mr2})
are known \cite{Kim:2005}:
\begin{subequations}
\eqnlab{eqm_coeffs}
\begin{align}
C^{(t)}_\perp=\frac{8(\lambda^2-1)}{3\lambda((2\lambda^2-3)\beta+1)}\,,\;\;\; &
C^{(t)}_\parallel=\frac{4(\lambda^2-1)}{3\lambda((2\lambda^2-1)\beta-1)}\,,\;\;\; \\
{K^{(r1)}_\perp=\frac{8a_\parallel a_\perp(\lambda^4-1)}{9\lambda^2 ((2\lambda^2-1)\beta-1)}\,,\;\;\; }&
{K^{(r1)}_\parallel=-\frac{8a_\parallel a_\perp(\lambda^2-1)}{9(\beta-1)\lambda^2}\,, }\\
{K^{(r2)}= -K^{(r1)}_{\perp} \frac{\lambda^2 - 1}{\lambda^2 + 1}\,, }\;\;\;  &
\beta=\frac{\ln[\lambda + \sqrt{\lambda^2-1}]}{\lambda\sqrt{\lambda^2-1}}\,.
\label{eq:beta}
\end{align}
\end{subequations}
Here $\lambda\obs{\equiv a_\parallel/as_\perp}$ is the aspect ratio of the spheroid, \obs{$2a_\parallel$ is the length of the particle symmetry axis, and $2a_\perp$ is  
the particle diameter.} Prolate spheroids
have $\lambda >1$. Oblate spheroids correspond to $\lambda < 1$.
In this case, Eq.~(\ref{eq:beta}) reduces to $\beta = \arccos(\lambda)/ [\lambda \sqrt{1 - \lambda^2}]$.  
The motion of the particle is fully characterised by the knowledge of $\ve v$,
the velocity of the particle, and the angular velocity of the particle
in the laboratory frame, $\ve \omega$. The direction of the particle symmetry vector, $\ve n$, is
related to the angular velocity $\ve \omega$ by the kinematic equation:
\begin{equation}
\frac{{\rm d} {\ve n} }{{\rm d}t } = {\ve \omega} \wedge {\ve n}\,.
\end{equation}
The equation for the angular velocity reads in the laboratory frame:
\begin{equation}
\label{eq:angvel}
\frac{{\rm d}}{{\rm d}t } \big(\ma J(\ve n)\ve \omega\big) = \ve T\,.
\end{equation}
Here $\ve T$ is the torque on the particle, and $\ma J(\ve n)$ is the inertia tensor of the particle.
For a spheroid its elements are given in Ref.~\cite{Kim:2005}.
A difficulty is that the inertia tensor depends on the instantaneous orientation $\ve n(t)$.
In the DNS we therefore express Eq.~(\ref{eq:angvel}) in the particle frame, at the cost of an extra
term in the equations of motion, due to the rotation of the
axes~\cite{LL_Mech}.

\section{Details on direct numerical simulations of turbulence}

\subsection{Method of resolution}
\label{subsec:num_resolution}
The Navier-Stokes equations read:
\begin{equation}
\rho_f[ \partial_t {\ve u} + ( {\ve u } \cdot \nabla ) {\ve u}] = - \nabla p + \nu\rho_{\rm f}\nabla^2 {\ve u} + {\ve \Phi}
\end{equation}
where $\ve u$ is the velocity field of the flow, which is assumed to be
incompressible:
\begin{equation}
\nabla \cdot {\ve u} = 0\,,
\end{equation}
$p$ is the pressure field, $\ve \Phi$ is a forcing term, $\rho_f$ is the density of the fluid, and $\nu$ is the kinematic viscosity.
The equations are solved in a triply periodic domain of
size $L^3$, using a pseudo-spectral method. The code has been described
elsewhere~\cite{Vos13}.
Briefly, the fluid is forced
in the band of lowest wave-numbers by imposing an energy injection of
$\varepsilon$.
The value of $\varepsilon$ chosen here correspond to cloud conditions, see
subsection~\ref{subsec:num_par}, in such a way that he Kolmogorov length,
$\eta_{\rm K} = (\nu^3/\varepsilon)^{1/4}$ is of the order of the grid spacing.
The code is fully dealiased, using the $2/3$-rule. Specifically, if $N$ is
the number of grid points in each direction, the nonlinear term is
computed by using only Fourier modes $0 \le n \le N/3$.

The equation of motion of the particles is implemented by interpolating
the velocity and velocity gradients at the location of the particle, using tri-cubic schemes. The
method has been carefully checked, by systematically comparing
the results in the absence of motion $\ve u = 0$, with the code, and with
an elementary solver (Mathematica) of the equations of motion of the particle.

\subsection{Choice of parameters }
\label{subsec:num_par}
The values chosen in this study correspond to cloud conditions~\cite{Pru78}.
In physical units, the viscosity chosen here is $\nu = 0.113 {\rm cm}^2/{\rm s}$.  The number of points taken here was $N=384$, with an energy injection rate of $\varepsilon \approx 1\,{\rm cm}^2/{\rm s}^3$, $N=784 $ ($\varepsilon \approx 16\,{\rm cm}^2/{\rm s}^3$), and $N = 1568 $ ($\varepsilon \approx 256\,{\rm cm}^2/{\rm s}^3$).  With the values of the parameters chosen here, the size $L$ of the box is equal to $8 \pi \,{\rm cm}$.

\section{Dependence of orientational bias on the Reynolds number}
\begin{figure}[t]
\begin{overpic}[width=5cm]{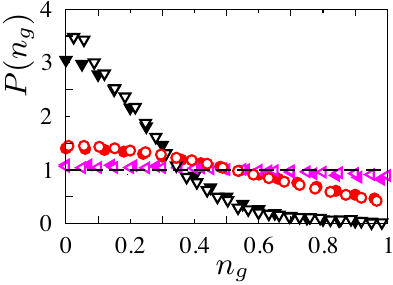}
\end{overpic}
\caption{\label{fig:comparison_SM}
Orientational bias of spheroids with aspect ratio $\lambda=1/50$ settling in turbulence.
Distribution $P(n_g)$ of $n_g\equiv|\ve n\cdot\gghat|$, where $\ve n$ is the particle symmetry-vector and $\hat{\ve g}$ is the direction of gravity.
DNS results for $P(n_g)$ with parameters ${\rm Re}_\lambda = 56$, $F_{\rm K}\approx 570$, ${\rm St}_{\rm K}\approx 0.02$ ($\blacktriangledown$); ${\rm Re}_\lambda = 95$, $F_{\rm K}\approx 70$, ${\rm St}_{\rm K}\approx 0.08$ ($\bullet$); and ${\rm Re}_\lambda = 151$, $F_{\rm K}\approx 9$, ${\rm St}_{\rm K}\approx 0.32$ ($\blacktriangleleft$).
Statistical-model simulations ($\ku=10$), as described in the text, open symbols.
The isotropic distribution $P(n_g)=1$ is shown as a dashed line.  } 
\end{figure}

\Figref{comparison_SM} shows a comparison between DNS simulations and the statistical model for the orientational bias in flows with different Reynolds numbers: ${\rm Re}_\lambda = 56$ ($\varepsilon=1\,  {\rm cm}^2 {\rm s}^{-3}$, $\blacktriangleleft$), ${\rm Re}_\lambda = 95$ ($\varepsilon=16\,{\rm cm}^2 {\rm s}^{-3}$, $\bullet$), ${\rm Re}_\lambda = 151$ ($\varepsilon=256\, {\rm cm}^2 {\rm s}^{-3}$, $\blacktriangledown$).
The DNS data is compared to statistical-model simulations with large $\ku$.
The statistical-model parameters $\st$, and $F$ are determined by the relations to the DNS parameters ${\rm Re}_\lambda$, $\st_{\rm K}$, and $F_{\rm K}$ described in the main text.
The agreement is equally good as that observed in Fig. 1 in the main text.

\section{Details on the statistical model}
In the simulations of the statistical model the smooth, incompressible, homogeneous, isotropic, threedimensional velocity field is given by $\ve u=\ve\nabla\wedge\ve\Psi/\sqrt{6}$, where $\ve\Psi$ is a vector potential~\cite{Gus16}.
Each component of $\ve\Psi$ consists of a spatial superposition of Fourier modes with random time-dependent prefactors.
The statistics of the prefactors is chosen such that $\Psi_i(\ve x,t)$ is Gaussian distributed with correlation function
\begin{align}
\langle\Psi_i(\ve x,t)\Psi_j(\ve x',t')\rangle=\ell^2u_0^2\exp\left[-\frac{|\ve x-\ve x'|^2}{2\ell^2}-\frac{|t-t'|}{\tau}\right]\,,
\eqnlab{corr_fun}
\end{align}
where $\langle\dots\rangle$ denotes an ensemble average.
We choose the velocity scale as the rms speed of the flow, $u_0\equiv\sqrt{\langle|\ve u|^2}\rangle$.
The correlation length $\ell$ and correlation time $\tau$ are the scales at which \Eqnref{corr_fun} decay in space and in time.
These Eulerian scales constitute the Kubo number $\ku=u_0\tau/\eta$~\cite{Gus16}.

\section{Details on the perturbation theory for small values of $\ku$}
\subsection{Implicit solution of equations of motion}
We follow the method outlined in Ref.~\cite{Gus16} and
expand the solution of the equations of motion for the settling spheroid, Eqs.~(1) and (2) in the main article, in powers of $\ku$:
\begin{align}
\begin{split}
\ve v(t)&=\sum_{n=0}^\infty (-1)^{n}\ku^n
\int_0^t{\rm d}t_1e^{\ma M^{(t)}_0(t_1-t)}\Delta\ma M^{(t)}(t_1)\cdots\int_0^{t_{n-1}}{\rm d}t_ne^{\ma M^{(t)}_0(t_n-t_{n-1})}\Delta\ma M^{(t)}(t_n)
\int_0^{t_{n}}{\rm d}t_{n+1}e^{\ma M^{(t)}_0(t_{n+1}-t_{n})}
\\&
\hspace{2cm}\times\big[{\rm F}\hat{\ve g}+\ma M^{(t)}(t_{n+1})\ve u(\ve x(t_{n+1}),t_{n+1})\big]\\
\ve \omega(t)&=\sum_{n=0}^\infty (-1)^{n}\ku^n
\int_0^t{\rm d}t_1e^{\ma M^{(r)}_0(t_1-t)}\Delta\ma M^{(r)}(t_1)\cdots\int_0^{t_{n-1}}{\rm d}t_ne^{\ma M^{(r)}_0(t_n-t_{n-1})}\Delta\ma M^{(r)}(t_n)
\int_0^{t_{n}}{\rm d}t_{n+1}e^{\ma M^{(r)}_0(t_{n+1}-t_{n})}
\\&
\hspace{2cm}\times\big[\tfrac{1}{2}\ma M^{(r)}(t_{n+1})\ve \Omega(\ve x(t_{n+1}),t_{n+1})-\Lambda C^{(r)}_\perp(\ma S(\ve x(t_{n+1}),t_{n+1})\ve n(t_{n+1}))\wedge\ve n(t_{n+1})
\\&
\hspace{4cm}+\ku\Lambda (\ve\omega(t_{n+1})\wedge\ve n(t_{n+1}))(\ve n(t_{n+1})\cdot\ve\omega(t_{n+1}))
\big]
\\
\ve n(t)&=\ve n_0+\ku\int_0^t{\rm d}t_1\ve\omega(t_1)\wedge\ve n(t_1)\,.
\end{split}
\eqnlab{eqm_sol_implicit}
\end{align}
Here $\Lambda \equiv (\lambda^2-1)/(\lambda^2+1)$, and we have decomposed the translational resistance tensor into
a part that does not depend on the Kubo number, and a $\ku$-dependent part,
 $\ma M^{(t)}=\ma M^{(t)}_0+\ku\Delta\ma M^{(t)}$, with
\begin{equation}
\ma M^{(t)}_0=\ma IC^{(t)}_\perp + \ve n_{0}\ve n_{0}\T(C^{(t)}_\parallel-C^{(t)}_\perp)\quad\mbox{and}\quad
 \Delta\ma M^{(t)}=[\Delta \ve n\,\ve n_0\T + \ve n_0\, \Delta \ve n\T + \ku \Delta \ve n\,\Delta \ve n\T](C^{(t)}_\parallel-C^{(t)}_\perp)\,.
\end{equation}
Here $\Delta\ve n$ is a rescaled orientation displacement vector
\begin{align}
\Delta\ve n(t)&\equiv\frac{\ve n(t)-\ve n_0}{\ku}=\int_0^t{\rm d}t_1\ve\omega(t_1)\wedge\ve n(t_1)\,.
\end{align}
The rotational resistance tensors are decomposed in a similar way.
\Eqnref{eqm_sol_implicit} is an exact, but implicit, solution to Eqs.~(1) and (2) in the main article,
provided that the initial conditions of the particle
velocity and angular velocity are set to zero.
The solution \eqnref{eqm_sol_implicit} depends implicitly on orientation $\ve n$,
the angular velocity $\ve\omega$, and on the centre-of-mass position $\ve x$ of the particle.
If the Kubo number is small we can terminate the solution \eqnref{eqm_sol_implicit} at some order in $\ku$ and solve the resulting set of equations iteratively. Details are given in Ref.~\cite{Gus16}.

\subsection{Explicit solution of equations of motion for small values of $\ku$}
To eliminate the implicit dependence on $\ve x$ through the flow velocity $\ve u(\ve x(t),t)$ and through the flow velocity gradients $\ma A(\ve x(t),t)$, we consider a decomposition of particle trajectories $\ve x(t)$
into two parts:
\begin{equation}
\ve x(t) \equiv \ve x^{({\rm d})}(t) + \delta\ve x(t)\,.
\end{equation}
The first part, $\ve x^{({\rm d})}(t)$,
is a deterministic part that
is  obtained by setting the turbulent velocity $\ve u$  to zero in the particle equation of motion.
The second part is the remainder, the flow-dependent fluctuating part, $\delta\ve x(t)\equiv\ve x(t) -\ve x^{({\rm d})}(t)$.
The deterministic part of the trajectory is obtained by integrating  the solution for the particle velocity in \Eqnref{eqm_sol_implicit} with $\ve u=0$ and $\ma A=0$.
For large values of $t$, the corresponding particle velocity approaches the settling velocity $\ve v_{\rm s}(\ve n_0)$ in a quiescent fluid
\begin{align}
\ve v_{\rm s}(\ve n_0)
& ={\rm F}\,\st\,[{\ma M^{(t)}(\ve n_0)}]^{-1}
\hat {\ve g}
  ={\rm F}\,\st \Big[\frac{\ma I}{C^{(t)}_\perp}+\frac{\ve n_0\ve n_0\T}{C^{(t)}_\parallel}-\frac{\ve n_0\ve n_0\T}{C^{(t)}_\perp}\Big]
\hat {\ve g}\,.
\end{align}
The corresponding deterministic trajectory approaches $\ve x^{({\rm d})}(t)
=\ve x_0+\ve v_{\rm s}(\ve n_0)t$.
Here $\ve n_0$ is the constant orientation at which the particle settles. In the steady state,
the orientation $\ve n_0$ must be chosen from a stationary distribution of particle orientations that must
be determined self consistently.

For small values of $\ku$, the flow gives rise to small deviations
 $\delta\ve x$ from the deterministic settling trajectories $\ve x^{({\rm d})}$.
Following the procedure outlined in Ref.~\cite{Gus16},
we attempt a series expansion of the fluid velocity experienced by particles in terms of small $\delta\ve x$
\begin{align}
\ve u(\ve x(t),t)=\ve u(\ve x^{({\rm d})}(t),t)+\delta\ve x\T(t)\ve\nabla\ve u(\ve x^{({\rm d})}(t),t)+\dots\,,
\end{align}
and similarly for the fluid-velocity gradients.
We insert these expansions into the implicit solutions \eqnref{eqm_sol_implicit}, and iterate the solution to find an expression for the orientation $\ve n$ valid for small values of $\ku$.
To second order in $\ku$ we find Eq. (3) in the main article.
Writing out  all terms explicitly this equation reads in vector notation:
\begin{align}
\ve n(t)&=
\ve n_0
+ \ku\int_0^t\!\!\!{\rm d}s\,[1-e^{(s-t)C^{(r)}_\perp/\st}]
\left.(\delta\ve x_{s}\cdot\ve\nabla)\big\{\ma O(\ve x,s)\ve n_0
+{\tfrac{\lambda^2-1}{\lambda^2+1}}\big[\ma S(\ve x,s)\ve n_0
- \big(\ve n_0\cdot \ma S(\ve x,s) \ve n_0\big)\ve n_0\big]\big\}\right|_{\ve x=\ve x_s^{(\rm d)}}
\nn\\&
+\ku\int_0^t{\rm d}s(1-e^{C^{(r)}_\perp(s-t)/\st})[\ma O_{s}\ve n_0+\Lambda(\ma S_{s}\ve n_0 - (\ve n_0\T\ma S_{s}\ve n_0)\ve n_0)]
\nn\\&
+\ku^2\int_0^t{\rm d}t_1\int_0^{t_1}{\rm d}t_2\Big[
d_1 \ve n_0\T\ma O_{t_1}\ma O_{t_2}
+ d_2 (\ve n_0\T\ma O_{t_1}\ma O_{t_2}\ve n_0) \ve n_0
+ d_3 (\ve n_0\T\ma O_{t_1}\ma S_{t_2}\ve n_0) \ve n_0
+ d_4 (\ve n_0\ma S_{t_1}\ve n_0)(\ve n_0\ma S_{t_2}\ve n_0)\ve n_0
\nn\\&
+ d_5 \ma O(t_2)\ma S(t_1)\ve n_0
+ d_6 (\ve n_0\T\ma S_{t_1}\ma O_{t_2}\ve n_0) \ve n_0
+ d_7 (\ve n_0\T\ma S_{t_1}\ma S_{t_2}\ve n_0) \ve n_0
+ d_8 (\ve n_0\ma S_{t_2}\ve n_0) \ma O_{t_1}\ve n_0
+ d_9 (\ve n_0\ma S_{t_1}\ve n_0) \ma O_{t_2}\ve n_0
\nn\\&
+ d_{10} \ma O_{t_1}\ma O_{t_2}\ve n_0
+ d_{11} \ma O_{t_1}\ma S_{t_2}\ve n_0
+ d_{12} (\ve n_0\ma S_{t_2}\ve n_0) \ma S_{t_1}\ve n_0
+ d_{13} (\ve n_0\ma S_{t_1}\ve n_0) \ma S_{t_2}\ve n_0
+ d_{14} \ma S_{t_1}\ma O_{t_2}\ve n_0
+ d_{15} \ma S_{t_1}\ma S_{t_2}\ve n_0
\Big]
\eqnlab{expansion_n}
\end{align}
with $\ma O_{t}\equiv\ma O(\ve x_t^{(\rm d)},t)$, $\ma S_{t}\equiv\ma S(\ve x_t^{(\rm d)},t)$.
The coefficents are:
\begin{align}
d_1&=-\frac{C^{(r)}_\perp (\Lambda -1) e^{C^{(r)}_\perp (t_1-t)/\st+C^{(r)}_\parallel (t_2-t)/\st}}{C^{(r)}_\perp+C^{(r)}_\parallel}+(\Lambda -1) e^{C^{(r)}_\perp (t_1-t)/\st+C^{(r)}_\parallel
   (t_2-t_1)/\st}-\frac{C^{(r)}_\parallel (\Lambda -1) e^{C^{(r)}_\parallel (t_2-t_1)/\st}}{C^{(r)}_\perp+C^{(r)}_\parallel}\nn\\
d_2&=(1-\Lambda ) e^{C^{(r)}_\perp (t_1-t)/\st+C^{(r)}_\parallel (t_2-t_1)/\st}\nn\\
&+\frac{C^{(r)}_\perp (\Lambda -1) e^{C^{(r)}_\perp (t_2-t)/\st+C^{(r)}_\parallel
   (t_1-t)/\st}}{C^{(r)}_\perp+C^{(r)}_\parallel}+\frac{C^{(r)}_\perp (\Lambda -1) e^{C^{(r)}_\perp (t_1-t)/\st+C^{(r)}_\parallel (t_2-t)/\st}}{C^{(r)}_\perp+C^{(r)}_\parallel}
\nn\\&
   +\frac{C^{(r)}_\parallel
   (\Lambda -1) e^{C^{(r)}_\parallel (t_2-t_1)/\st}}{C^{(r)}_\perp+C^{(r)}_\parallel}+\frac{(C^{(r)}_\perp+C^{(r)}_\parallel \Lambda ) e^{C^{(r)}_\perp
   (t_2-t_1)/\st}}{C^{(r)}_\perp+C^{(r)}_\parallel}+e^{C^{(r)}_\perp (-2 t+t_1+t_2)/\st}+(-\Lambda -1) e^{C^{(r)}_\perp (t_2-t)/\st}\nn\\
d_3&=\frac{C^{(r)}_\perp (\Lambda -1) \Lambda  e^{C^{(r)}_\perp (t_2-t)/\st+C^{(r)}_\parallel (t_1-t)/\st}}{C^{(r)}_\perp+C^{(r)}_\parallel}+\frac{\Lambda  (C^{(r)}_\perp+C^{(r)}_\parallel \Lambda )
   e^{C^{(r)}_\perp (t_2-t_1)/\st}}{C^{(r)}_\perp+C^{(r)}_\parallel}\nn\\
&+\Lambda  e^{C^{(r)}_\perp (-2 t+t_1+t_2)/\st}-(\Lambda +1) \Lambda  e^{C^{(r)}_\perp (t_2-t)/\st}\nn\\
d_4&=-3 \Lambda ^2 e^{C^{(r)}_\perp (t_1-t)/\st}+\Lambda ^2 e^{C^{(r)}_\perp (-2 t+t_1+t_2)/\st}+\Lambda ^2 e^{C^{(r)}_\perp (t_2-t)/\st}-2 \Lambda ^2 e^{C^{(r)}_\perp(t_2-t_1)/\st}+3 \Lambda ^2\nn\\
d_5&=(\Lambda -1) \Lambda  e^{C^{(r)}_\perp (t_1-t)/\st+C^{(r)}_\parallel (t_2-t_1)/\st}-\frac{C^{(r)}_\parallel (\Lambda -1) \Lambda  e^{C^{(r)}_\parallel
   (t_2-t_1)/\st}}{C^{(r)}_\perp+C^{(r)}_\parallel}\nn
\end{align}
\begin{align}
d_6&=-\frac{C^{(r)}_\perp (\Lambda -1) \Lambda  e^{C^{(r)}_\perp (t_1-t)/\st+C^{(r)}_\parallel (t_2-t)/\st}}{C^{(r)}_\perp+C^{(r)}_\parallel}+(\Lambda -1) \Lambda  e^{C^{(r)}_\perp
   (t_1-t)/\st+C^{(r)}_\parallel (t_2-t_1)/\st}-\frac{C^{(r)}_\parallel (\Lambda -1) \Lambda  e^{C^{(r)}_\parallel (t_2-t_1)/\st}}{C^{(r)}_\perp+C^{(r)}_\parallel}
\nn\\&
   +2 \Lambda
   e^{C^{(r)}_\perp (t_1-t)/\st}-\Lambda  e^{C^{(r)}_\perp (-2 t+t_1+t_2)/\st}+\Lambda  e^{C^{(r)}_\perp (t_2-t_1)/\st}-2 \Lambda\nn\\
d_7&=2 \Lambda ^2 e^{C^{(r)}_\perp (t_1-t)/\st}-\Lambda ^2 e^{C^{(r)}_\perp (-2 t+t_1+t_2)/\st}+\Lambda ^2 e^{C^{(r)}_\perp (t_2-t_1)/\st}-2 \Lambda ^2\nn\\
d_8&=\frac{C^{(r)}_\perp (\Lambda -1) \Lambda  e^{C^{(r)}_\perp (t_2-t)/\st+C^{(r)}_\parallel (t_1-t)/\st}}{C^{(r)}_\perp+C^{(r)}_\parallel}+\frac{\Lambda  (C^{(r)}_\perp+C^{(r)}_\parallel \Lambda )
   e^{C^{(r)}_\perp (t_2-t_1)/\st}}{C^{(r)}_\perp+C^{(r)}_\parallel}\nn\\
&+\Lambda  e^{C^{(r)}_\perp (t_1-t)/\st}+\Lambda ^2 \left(-e^{C^{(r)}_\perp (t_2-t)/\st}\right)-\Lambda\nn\\
d_9&=\frac{C^{(r)}_\perp (\Lambda -1) \Lambda  e^{C^{(r)}_\perp (t_1-t)/\st+C^{(r)}_\parallel (t_2-t)/\st}}{C^{(r)}_\perp+C^{(r)}_\parallel}-(\Lambda -1) \Lambda  e^{C^{(r)}_\perp
   (t_1-t)/\st+C^{(r)}_\parallel (t_2-t_1)/\st}+\frac{C^{(r)}_\parallel (\Lambda -1) \Lambda  e^{C^{(r)}_\parallel (t_2-t_1)/\st}}{C^{(r)}_\perp+C^{(r)}_\parallel}
\nn\\&
   +\Lambda
   e^{C^{(r)}_\perp (t_1-t)/\st}-\Lambda  e^{C^{(r)}_\perp (t_2-t)/\st}+\Lambda  e^{C^{(r)}_\perp (t_2-t_1)/\st}-\Lambda\nn\\
d_{10}&=\frac{(C^{(r)}_\perp-C^{(r)}_\perp \Lambda ) e^{C^{(r)}_\perp (t_2-t)/\st+C^{(r)}_\parallel (t_1-t)/\st}}{C^{(r)}_\perp+C^{(r)}_\parallel}-\frac{(C^{(r)}_\perp+C^{(r)}_\parallel \Lambda )
   e^{C^{(r)}_\perp (t_2-t_1)/\st}}{C^{(r)}_\perp+C^{(r)}_\parallel}-e^{C^{(r)}_\perp (t_1-t)/\st}+\Lambda  e^{C^{(r)}_\perp (t_2-t)/\st}+1\nn\\
d_{11}&=-\frac{C^{(r)}_\perp (\Lambda -1) \Lambda  e^{C^{(r)}_\perp (t_2-t)/\st+C^{(r)}_\parallel (t_1-t)/\st}}{C^{(r)}_\perp+C^{(r)}_\parallel}-\frac{\Lambda  (C^{(r)}_\perp+C^{(r)}_\parallel \Lambda )
   e^{C^{(r)}_\perp (t_2-t_1)/\st}}{C^{(r)}_\perp+C^{(r)}_\parallel}\nn\\
&-\Lambda  e^{C^{(r)}_\perp (t_1-t)/\st}+\Lambda ^2 e^{C^{(r)}_\perp (t_2-t)/\st}+\Lambda\nn\\
d_{12}&=\Lambda ^2 e^{C^{(r)}_\perp (t_1-t)/\st}-\Lambda ^2 e^{C^{(r)}_\perp (t_2-t)/\st}+\Lambda ^2 e^{C^{(r)}_\perp (t_2-t_1)/\st}-\Lambda ^2\nn\\
d_{13}&=\Lambda ^2 e^{C^{(r)}_\perp (t_1-t)/\st}-\Lambda ^2 e^{C^{(r)}_\perp (t_2-t)/\st}+\Lambda ^2 e^{C^{(r)}_\perp (t_2-t_1)/\st}-\Lambda ^2\nn\\
d_{14}&=\frac{C^{(r)}_\perp (\Lambda -1) \Lambda  e^{C^{(r)}_\perp (t_1-t)/\st+C^{(r)}_\parallel (t_2-t)/\st}}{C^{(r)}_\perp+C^{(r)}_\parallel}+\Lambda  \left(-e^{C^{(r)}_\perp
   (t_1-t)/\st}\right)+\Lambda  e^{C^{(r)}_\perp (t_2-t)/\st}-\Lambda  e^{C^{(r)}_\perp (t_2-t_1)/\st}+\Lambda\nn\\
d_{15}&=\Lambda ^2 \left(-e^{C^{(r)}_\perp (t_1-t)/\st}\right)+\Lambda ^2 e^{C^{(r)}_\perp (t_2-t)/\st}-\Lambda ^2 e^{C^{(r)}_\perp (t_2-t_1)/\st}+\Lambda ^2
\nn
\end{align}
where $\Lambda=(\lambda^2-1)/(\lambda^2+1)$.
To obtain the first term in the last row of Eq.~(3) in the main text, for example, we must
add the terms involving $d_1$, $d_2$, and $d_{10}$.

\subsection{Moments of alignment}
Eq.~\eqnref{expansion_n} relates the orientation dynamics to the fluid velocity and its gradients.
This expression makes it possible to calculate the moments $\langle(\ve n\cdot\hat{\ve g})^{2p}\rangle_{\ve n_0}$ conditional on $\ve n_0$ (odd-order moments vanish because spheroids are fore-aft symmetric).
By raising \Eqnref{expansion_n} to the power $2p$, terminating the resulting expressions at order $\ku^2$, and taking the average using the known stationary statistics of the fluid velocity and its gradients,
we obtain the moments of $\ve n\cdot\hat{\ve g}$ conditional on initial alignment $\ve n_0\cdot\hat{\ve g}$.

However, the resulting expressions contain secular terms. These terms are proportional to $\ku^2$ and grow linearly with time.
We know that such secular terms must vanish, because the vector $\ve n$ must remain normalised to unity.
Its components cannot continue to grow.
We therefore demand that the
initial alignment $\ve n_0\cdot\hat{\ve g}$ is distributed
according to a stationary distribution of alignments that
causes all secular terms to vanish.
This procedure gives rise to one condition per value of $p$, and this
 is sufficient to determine the desired stationary distribution of alignments.

The dependence of the secular terms upon $(\ve n_0\cdot\hat{\ve g})^{2p}$ is not only through algebraic powers, but also functional through exponentials and error functions. As a consequence, the conditions that make sure that the secular terms vanish are hard to solve in general. We therefore seek a perturbative solution and expand
the secular conditions and the moments
$m_p\equiv\langle {(\ve n_0\cdot\hat{\ve g} )^{2p}\rangle}$ in terms of
the parameter $G\equiv\ku F/C^{(t)}_\perp$, for instance:
\begin{equation}
 m_p=\sum_{i=0}^\infty m_p^{(i)}G^i\,.
\end{equation}
To lowest order in $G$ we find the recursion
\begin{align}
(1 + 2 p) m^{(0)}_p + (1 - 2 p) m^{(0)}_{p-1} = 0\,.
\end{align}
Using the boundary condition $m^{(0)}_{0}=1$ we find
\begin{align}
m^{(0)}_p=\frac{1}{2p+1}\,.
\end{align}
These moments correspond to a uniform distribution
of $\ve n_0$.
Higher-order contributions in $\G$ to the moments can be recursively
determined by series expansion of the secular terms.
We find that moments $m^{(i)}_p$ with odd values of $i$ vanish,
 and that moments $m^{(i)}_p$ with even values of $i$
are on the form given by Eq. (6) in the main article:
\begin{align}
\langle(\ve n\cdot\gghat)^{2p}\rangle_\infty\!=\!
\frac{1}{2p\!+\!1}\!+\!
\sum_{i=1}^\infty\frac{\G^{2i}\sum_{j=1}^i p^jA^{(2i)}_{j}(\st,\lambda)}{\prod_{k=1}^{i+1}(2 p + 2 k - 1)}\,.
\end{align}
The coefficients $A^{(2i)}_{j}(\st,\lambda)$ are quite lengthy in general.
We therefore only give the lowest-order expression here.
To order $\G^2$ we have
\begin{align}
\langle(\ve n\cdot\gghat)^{2p}\rangle_\infty\!=\!
\frac{1}{2p\!+\!1}\!+\!
\frac{\G^{2}pA^{(2)}_{1}(\st,\lambda)}{(2p+1)(2p+3)}\,.
\end{align}
Decomposing $A^{(2)}_{1}(\st,\lambda)$ in one contribution due to preferential sampling and one due to the history effect, we have
\begin{align}
A^{(2)}_{1}(\st,\lambda)=A^{(2)}_{1,{\rm pref.}}(\st,\lambda)+A^{(2)}_{1,{\rm hist.}}(\st,\lambda)\,.
\eqnlab{coeffA21}
\end{align}
The preferential-sampling contribution reads:
\begin{align}
A^{(2)}_{1,{\rm pref.}}(\st,\lambda)=\frac{7-\Lambda}{5+3\Lambda^2}\frac{4C^{(t)}_\perp\st^3(C^{(t)}_\perp-C^{(t)}_\parallel) \Big[\big(C^{(t)}_\perp\big)^2+\big(C^{(t)}_\parallel\big)^2+C^{(t)}_\perp C^{(t)}_\parallel
+3\st\big(C^{(t)}_\perp+C^{(t)}_\parallel+3\st\big)\Big]}{3 (C^{(t)}_\perp+\st)^3 C^{(t)}_\parallel (C^{(t)}_\parallel+\st)^3}\,.
\eqnlab{coeffA21preferential}
\end{align}
The contribution due to the history effect is
\begin{align}
&A^{(2)}_{1,{\rm history}}(\st,\lambda)=
\frac{1}{3 {C^{(t)}_\parallel}^2 \left(3 \Lambda ^2+5\right) (10 {C^{(t)}_\parallel}+3 {\st})^3 (10 {C^{(t)}_\perp} {C^{(t)}_\parallel}-3 {C^{(t)}_\perp} {\st}+6 {C^{(t)}_\parallel}
   {\st})^3}
\Big[
\nn\\&
-2000000\, {C^{(t)}_\perp}^3 {C^{(t)}_\parallel}^6 \left(-4 {C^{(t)}_\perp}^2 \left(4 \Lambda ^2+7 \Lambda +7\right)+{C^{(t)}_\perp} {C^{(t)}_\parallel} \left(3
   \Lambda ^2+28 \Lambda -7\right)+{C^{(t)}_\parallel}^2 \left(19 \Lambda ^2-42 \Lambda +35\right)\right)
\nn\\&
-3600000\,{\st} {C^{(t)}_\perp}^2 {C^{(t)}_\parallel}^6
    \left(-4 {C^{(t)}_\perp}^2 \left(4 \Lambda ^2+7 \Lambda +7\right)+{C^{(t)}_\perp} {C^{(t)}_\parallel} \left(3 \Lambda ^2+28 \Lambda -7\right)+{C^{(t)}_\parallel}^2 \left(19
   \Lambda ^2-42 \Lambda +35\right)\right)
\nn\\&
-540000\,{\st}^2 {C^{(t)}_\perp} {C^{(t)}_\parallel}^4  \left({C^{(t)}_\perp}^2-2 {C^{(t)}_\perp} {C^{(t)}_\parallel}-4 {C^{(t)}_\parallel}^2\right) \Big(4
   {C^{(t)}_\perp}^2 \left(4 \Lambda ^2+7 \Lambda +7\right)+{C^{(t)}_\perp} {C^{(t)}_\parallel} \left(-3 \Lambda ^2-28 \Lambda +7\right)
\nn\\&\hspace{12.9cm}
   +{C^{(t)}_\parallel}^2 \left(-19 \Lambda ^2+42
   \Lambda -35\right)\Big)
\nn\\&
-108000\,{\st}^3 {C^{(t)}_\parallel}^4  \Big(7 {C^{(t)}_\perp}^4 \left(15
   \Lambda ^2+22 \Lambda +23\right)-21 {C^{(t)}_\perp}^3 {C^{(t)}_\parallel} \left(11 \Lambda ^2+24 \Lambda +13\right)-4 {C^{(t)}_\perp}^2 {C^{(t)}_\parallel}^2 \left(32 \Lambda ^2-119
   \Lambda +105\right)
\nn\\&\hspace{8cm}
   +8 {C^{(t)}_\perp} {C^{(t)}_\parallel}^3 \left(30 \Lambda ^2-49 \Lambda +49\right)+4 {C^{(t)}_\parallel}^4 \left(19 \Lambda ^2-42 \Lambda
   +35\Big)\right)
\nn\\&
-48600\,{\st}^4 {C^{(t)}_\parallel}^2  \Big({C^{(t)}_\perp}^5 \left(-\left(23 \Lambda ^2+28 \Lambda +21\right)\right)+14 {C^{(t)}_\perp}^4 {C^{(t)}_\parallel}
   \left(7 \Lambda ^2+12 \Lambda +5\right)+{C^{(t)}_\perp}^3 {C^{(t)}_\parallel}^2 \left(-41 \Lambda ^2-210 \Lambda +119\right)
\nn\\&\hspace{3.5cm}
   -8 {C^{(t)}_\perp}^2 {C^{(t)}_\parallel}^3 \left(22 \Lambda
   ^2+7 \Lambda +49\right)+8 {C^{(t)}_\perp} {C^{(t)}_\parallel}^4 \left(3 \Lambda ^2+28 \Lambda -7\right)+8 {C^{(t)}_\parallel}^5 \left(19 \Lambda ^2-42 \Lambda
   +35\right)\Big)
\nn\\&
+29160\,{\st}^5 {C^{(t)}_\parallel}^2  ({C^{(t)}_\perp}-2 {C^{(t)}_\parallel})^2 \left(14 {C^{(t)}_\perp}^2 \left(\Lambda ^2+3 \Lambda +2\right)+{C^{(t)}_\perp}
   {C^{(t)}_\parallel} \left(-3 \Lambda ^2-28 \Lambda +7\right)+{C^{(t)}_\parallel}^2 \left(-19 \Lambda ^2+42 \Lambda -35\right)\right)
\nn\\&
-1458\, {\st}^6 ({C^{(t)}_\perp}-2 {C^{(t)}_\parallel})^3 \left({C^{(t)}_\perp}^2 \left(9 \Lambda ^2+28 \Lambda +35\right)+{C^{(t)}_\parallel}^2 \left(-19 \Lambda ^2+42 \Lambda
   -35\right)\right)
\Big]\,.
\eqnlab{coeffA21history}
\end{align}
Here we have used the identities
\begin{align}
C^{(r)}_\perp &= \frac{10}{3}C^{(t)}_\parallel \quad\mbox{and}\quad  C^{(r)}_\parallel = \frac{10}{3}\frac{C^{(t)}_\perp C^{(t)}_\parallel}{2C^{(t)}_\parallel - C^{(t)}_\perp}
\end{align}
to simplify.
\Eqnref{eqm_coeffs} relates $C^{(t)}_\perp$ and $C^{(t)}_\parallel$ to $\lambda$, so that $A^{(2)}_{1}$ is a function of $\st$ and $\lambda$ only.
\Figref{coeffA21} shows how the expressions for $A^{(2)}_{1}$ and for the relative preferential-sampling contribution $|A^{(2)}_{1,{\rm pref.}}/A^{(2)}_{1}|$ depend on the aspect
ratio  $\lambda$, for different values of $\st$. We see that the preferential contribution is small compared to the contribution due to the history effect for all parameter values.

\begin{figure}
\begin{overpic}{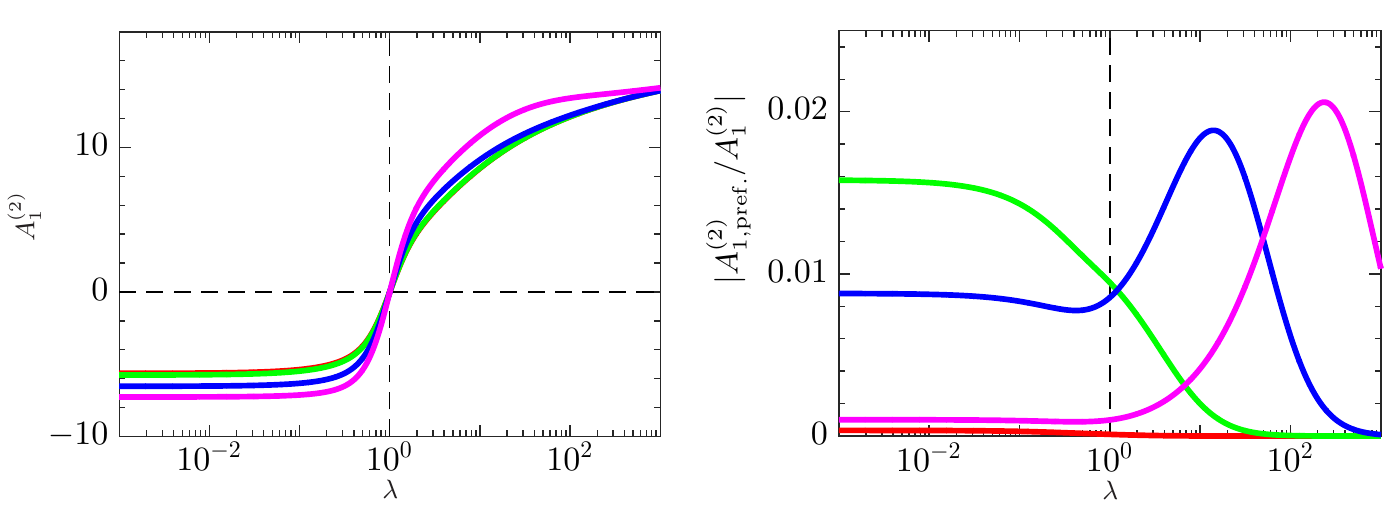}
\end{overpic}
\caption{Left: Coefficient $A^{(2)}_{1}(\st,\lambda)$ [Eqs. \eqnref{coeffA21}--\eqnref{coeffA21history}] plotted against $\lambda$ for $\st=0.1$ (red), $1$ (green), $10$ (blue), $100$ (magenta).
 Right: Corresponding plot for the relative contribution due to preferential sampling $|A^{(2)}_{1,{\rm pref.}}(\st,\lambda)/A^{(2)}_{1}(\st,\lambda)|$.}
\figlab{coeffA21}
\end{figure}

\subsection{\em Large-$\rm G$ limit}
In the previous Section we summarised how the moments of the orientation distribution
can be calculated by using perturbation
expansions in the parameter ${\rm G}$. Now consider the opposite limit of large values of $\G$. In this limit
we obtain a recursion equation for the moments $\langle(\ve n\cdot\hat{\ve g})^{2p}\rangle$ that is independent of preferential effects.
This is expected because the particles fall rapidly through the turbulence in this limit, experiencing the turbulent gradients
approximately as a white-noise signal. The history contribution is obtaind from a fourth-order recursion that is hard to solve in general.

One might expect that the orientation of rod-like particles approaches perfect alignment with gravity, that
is $\langle(\ve n\cdot\hat{\ve g})^{2p}\rangle=1$ for all values of $p$ as $G\to\infty$.
However, when we insert this limiting expression into the fourth-order recursion equation, we obtain a condition that is only satisfied when $\lambda=\infty$.
This shows that perfect alignment is only possible when $\lambda=\infty$.
For other values of $\lambda$ the moments must approach limiting values as $G\to\infty$. We have not yet managed to calculate these limits from our statistical-model theory.

